\documentclass[aip,jcp,amsmath,amssymb,preprint]{revtex4-1}

\usepackage{dcolumn} 
\usepackage{bm}
\usepackage{graphicx}
\usepackage{longtable}

\usepackage{color}

\draft 

\begin{document}

\newcommand{\bra}[1]{\langle #1|}
\newcommand{\ket}[1]{|#1\rangle}
\newcommand{\braket}[2]{\langle #1|#2\rangle}

\title{Accurate \textit{ab initio} vibrational energies of methyl chloride}

\author{Alec Owens} 
\email{owens@mpi-muelheim.mpg.de}
\affiliation{Max-Planck-Institut f\"{u}r Kohlenforschung, Kaiser-Wilhelm-Platz 1, 45470 M\"{u}lheim an der Ruhr, Germany}
\affiliation{Department of Physics and Astronomy, University College London, Gower Street, WC1E 6BT London, United Kingdom}

\author{Sergei N. Yurchenko}
\affiliation{Department of Physics and Astronomy, University College London, Gower Street, WC1E 6BT London, United Kingdom}

\author{Andrey Yachmenev}
\affiliation{Department of Physics and Astronomy, University College London, Gower Street, WC1E 6BT London, United Kingdom}

\author{Jonathan Tennyson}
\affiliation{Department of Physics and Astronomy, University College London, Gower Street, WC1E 6BT London, United Kingdom}

\author{Walter Thiel}
\affiliation{Max-Planck-Institut f\"{u}r Kohlenforschung, Kaiser-Wilhelm-Platz 1, 45470 M\"{u}lheim an der Ruhr, Germany}

\date{\today}

\begin{abstract}
 Two new nine-dimensional potential energy surfaces (PESs) have been generated using high-level \textit{ab initio} theory for the two main isotopologues of methyl chloride, CH$_{3}{}^{35}$Cl and CH$_{3}{}^{37}$Cl. The respective PESs, CBS-35$^{\,\mathrm{HL}}$ and CBS-37$^{\,\mathrm{HL}}$, are based on explicitly correlated coupled cluster calculations with extrapolation to the complete basis set (CBS) limit, and incorporate a range of higher-level (HL) additive energy corrections to account for core-valence electron correlation, higher-order coupled cluster terms, scalar relativistic effects, and diagonal Born-Oppenheimer corrections. Variational calculations of the vibrational energy levels were performed using the computer program TROVE, whose functionality has been extended to handle molecules of the form XY$_3$Z. Fully converged energies were obtained by means of a complete vibrational basis set extrapolation. The CBS-35$^{\,\mathrm{HL}}$ and CBS-37$^{\,\mathrm{HL}}$ PESs reproduce the fundamental term values with root-mean-square errors of $0.75$ and $1.00{\,}$cm$^{-1}$ respectively. An analysis of the combined effect of the HL corrections and CBS extrapolation on the vibrational wavenumbers indicates that both are needed to compute accurate theoretical results for methyl chloride. We believe that it would be extremely challenging to go beyond the accuracy currently achieved for CH$_3$Cl without empirical refinement of the respective PESs.
\end{abstract}


\maketitle 

\section{Introduction}
\label{sec:intro}
 
 Methyl chloride has been proposed as an observable biosignature gas in the search for life outside the Solar System.~\cite{05SeKaMe.CH3Cl} A model of different hypothetical Earth-like planets orbiting a range of M stars predicted that a higher concentration of CH$_{3}$Cl would exist than on Earth, and with stronger spectral features. Seager et al. have since gone on to classify CH$_3$Cl as a Type III biomarker - a molecule produced from a secondary metabolic process - and estimated the required concentration needed for a realistic detection in a generalized oxidized atmosphere,~\cite{13SeBaHu.CH3Cl} and for an exoplanet with a thin H$_2$ rich atmosphere and a habitable surface temperature.~\cite{13aSeBaHu.CH3Cl} The rotation-vibration spectrum of CH$_3$Cl has received increased interest as a result.
 
  A highly accurate and comprehensive line list is lacking for methyl chloride, with varying coverage in the spectroscopic databases.~\cite{HITRAN,PNNL,GEISA,JPL} The HITRAN database~\cite{HITRAN} is the most extensive, containing over $100{\,}000$ transitions for each of the two main isotopologues, $^{12}$CH$_{3}{}^{35}$Cl and $^{12}$CH$_{3}{}^{37}$Cl (henceforth labelled as CH$_{3}{}^{35}$Cl and CH$_{3}{}^{37}$Cl), in the range $0$ to $3200{\,}$cm$^{-1}$. Although the latest update HITRAN2012 has seen improvements, notably around $3000{\,}$cm$^{-1}$,~\cite{11BrPeJa.CH3Cl} there are still deficiencies with certain line positions and intensities taken from an empirically refined theoretical anharmonic force field.~\cite{85KoKoNa.CH3Cl}
 
  Due to its prominent role in depletion of the ozone layer, levels of methyl chloride are being closely monitored by satellite missions such as the Atmospheric Chemistry Experiment~\cite{ACE_Bern06,06NaBeBo.CH3Cl,11BrChMa.CH3Cl,13BrVoSc.CH3Cl} and the Microwave Limb Sounder.~\cite{13SaLiMa.CH3Cl} A number of recent publications focusing on line shapes and broadening coefficients,~\cite{12BrJaBu.CH3Cl,13BrJaLa.CH3Cl,13aBrJaLa.CH3Cl,12GuRoBu.CH3Cl,
  11BuGuEl.CH3Cl,12BuRoxx.CH3Cl,13Buxxxx.CH3Cl,13BuMaMo.CH3Cl,
  13RaJaDh.CH3Cl,14RaJaDh.CH3Cl,14aRaJaDh.CH3Cl,13DuLaBu.CH3Cl,14DuLaBu.CH3Cl} needed for a realistic modelling of atmospheric spectra, confirm its terrestrial importance. The $3.4{\,}\mu$m region is particularly relevant for atmospheric remote sensing due to a relatively transparent window and strong spectral features of the $\nu_1$ band of CH$_{3}$Cl. A high-resolution study of the $\nu_1$, $\nu_4$ and $3\nu_6$ bands in this region produced a line list for the range $2920$ to $3100{\,}$cm$^{-1}$.~\cite{11BrPeJa.CH3Cl} The $6.9{\,}\mu$m region has seen line positions, intensities, and self-broadening coefficients determined for more than 900 rovibrational transitions of the $\nu_{5}$ band.~\cite{13RaJaDh.CH3Cl} Nikitin et al. have also measured, modelled and assigned over $20{\,}000$ transitions for each isotopologue in the region $0$ to $2600{\,}$cm$^{-1}$.~\cite{03NiFeCh.CH3Cl,04NiChBu.CH3Cl,05NiChxx.CH3Cl,05NiChBu.CH3Cl} An effective Hamiltonian model adapted to the polyad structure of methyl chloride reproduced observed transitions involving the ground state and $13$ vibrational states with an overall standard deviation of $0.0003{\,}$cm$^{-1}$.
 
 There is a large body of experimental work on the rovibrational spectrum of methyl chloride. We refer the reader to the most recent publications~\cite{11BrPeJa.CH3Cl,12BrTrJa.CH3Cl,12BrJaBu.CH3Cl,13BrJaLa.CH3Cl,13aBrJaLa.CH3Cl,12GuRoBu.CH3Cl,11BuGuEl.CH3Cl,12BuRoxx.CH3Cl,13Buxxxx.CH3Cl,13BuMaMo.CH3Cl,13RaJaDh.CH3Cl,14RaJaDh.CH3Cl,14aRaJaDh.CH3Cl,13DuLaBu.CH3Cl,14DuLaBu.CH3Cl,03NiFeCh.CH3Cl,04NiChBu.CH3Cl,05NiChxx.CH3Cl,05NiChBu.CH3Cl} (and references therein) for a more complete overview.
 
 Theoretically there has been a consistent effort over the years to characterize the spectrum of CH$_3$Cl. Much attention has been given to a description of harmonic~\cite{70DuAlMc.CH3Cl,72DuMcSp.CH3Cl,76Duxxxx.CH3Cl,79Wixxxx.CH3Cl,80LaTaBe.CH3Cl,87ScThxx.CH3Cl,95Haxxxx.CH3Cl,01BlLaxx.CH3Cl} and anharmonic~\cite{77ScWoBe.CH3Cl,83BeAlxx.CH3Cl,83KoKoNa.CH3Cl,84KoKoNa.CH3Cl,85KoKoNa.CH3Cl,90DuLaxx.CH3Cl,92ScThxx.CH3Cl} force fields, both empirically and using \textit{ab initio} methods. The latest work by Black and Law~\cite{01BlLaxx.CH3Cl} employed spectroscopic data from ten isotopomers of methyl chloride to produce an empirical harmonic force field incorporating the most up to date treatment of anharmonic corrections. These were largely based on a complete set of empirical anharmonicity constants derived from a joint local mode and normal mode analysis of 66 vibrational energy levels in the region $700$ to $16{\,}500{\,}$cm$^{-1}$,~\cite{90DuLaxx.CH3Cl} and follow-up work in a similar vein by Law.~\cite{99Laxxxx.CH3Cl}
  
  From a purely \textit{ab initio} standpoint, Nikitin has computed global nine-dimensional potential energy surfaces (PESs) for vibrational energy level calculations, considering both CH$_{3}{}^{35}$Cl and CH$_{3}{}^{37}$Cl in the region $0$ to $3500{\,}$cm$^{-1}$.~\cite{08Nixxxx.CH3Cl} Using fourth-order M{\o}ller-Plesset perturbation theory MP4 and a correlation-consistent quadruple-zeta basis set, as well as coupled cluster theory CCSD(T) with a triple-zeta basis set, a combined total of $7241$ points with energies up to $h c \cdot 40{\,}000{\,}$cm$^{-1}$ were employed to generate and fit the PESs ($h$ is the Planck constant and $c$ is the speed of light). Vibrational energies were calculated variationally using a finite basis representation and an exact kinetic energy operator, reproducing the fundamental term values with a root-mean-square error of $1.97$ and $1.71{\,}$cm$^{-1}$ for CH$_{3}{}^{35}$Cl and CH$_{3}{}^{37}$Cl respectively.

 The potential energy surface is the foundation of rovibrational energy level calculations. Its quality not only dictates the accuracy of line positions, but is also crucial for achieving significant improvements in calculated band intensities.~\cite{OvThYu08a.PH3} Achieving ``spectroscopic accuracy'' (better than $\pm 1{\,}$cm$^{-1}$) in a purely \textit{ab initio} fashion is extremely challenging due to the limitations of electronic structure methods. To do so one must account for higher-level (HL) electron correlation beyond the initial coupled cluster method when generating the PES, and use a one-particle basis set near the complete basis set (CBS) limit. Core-valence (CV) electron correlation, scalar relativistic (SR) effects, higher-order (HO) electron correlation, and the diagonal Born-Oppenheimer correction (DBOC) are considered to be the leading HL contributions.~\cite{Peterson12}
 
 The goal of this study is to use state-of-the-art electronic structure calculations to construct a global PES for each of the two main isotopologues of methyl chloride, CH$_{3}{}^{35}$Cl and CH$_{3}{}^{37}$Cl. This requires inclusion of the leading HL corrections and extrapolation to the CBS limit. The quality of the respective PESs will be assessed by variational calculations of the vibrational energy levels using the computer program TROVE.~\cite{TROVE2007} To obtain fully converged term values we will exploit the smooth convergence of computed energies with respect to vibrational basis set size, and perform a complete vibrational basis set (CVBS) extrapolation.~\cite{OvThYu08.PH3} By using a range of theoretical techniques we aim to find out exactly what accuracy is possible for a molecule such as methyl chloride.

 The paper is structured as follows: In Sec.~\ref{sec:PES} the \textit{ab initio} calculations and analytic representation of the PES are detailed. The variational calculations will be discussed in Sec.~\ref{sec:variational}, where we assess the combined effect of the HL corrections and CBS extrapolation on the vibrational term values and equilibrium geometry. We offer concluding remarks in Sec.~\ref{sec:conc}.
 
\section{Potential Energy Surface}
\label{sec:PES} 

\subsection{Electronic structure calculations}
 
 We take a focal-point approach~\cite{Csaszar98} to represent the total electronic energy,
\begin{equation}\label{eq:tot_en}
E_{\mathrm{tot}} = E_{\mathrm{CBS}}+\Delta E_{\mathrm{CV}}+\Delta E_{\mathrm{HO}}+\Delta E_{\mathrm{SR}}+\Delta E_{\mathrm{DBOC}}
\end{equation}

\noindent which allows for greater control over the PES. To compute $E_{\mathrm{CBS}}$ we employed the explicitly correlated F12 coupled cluster method CCSD(T)-F12b~(Ref.~\onlinecite{Adler07} - for a detailed review of this method see Refs.~\onlinecite{F12_Tew2010,F12_Werner2010}) in conjunction with the F12-optimized correlation consistent polarized valence basis sets, cc-pVTZ-F12 and cc-pVQZ-F12,~\cite{Peterson08} in the frozen core approximation. The diagonal fixed amplitude ansatz 3C(FIX),~\cite{TenNo04} and a Slater geminal exponent value of $\beta=1.0$~$a_0^{-1}$ as recommended by Hill et al.~\cite{Hill09} were used. To evaluate the many electron integrals in F12 theory additional auxiliary basis sets (ABS) are required. For the resolution of the identity (RI) basis, and the two density fitting (DF) basis sets, we utilized the corresponding OptRI,~\cite{Yousaf08} cc-pV5Z/JKFIT,~\cite{Weigend02} and aug-cc-pwV5Z/MP2FIT~\cite{Hattig05} ABS, respectively. All calculations were carried out using MOLPRO2012~\cite{Werner2012} unless stated otherwise.
 
 To extrapolate to the CBS limit we used a parameterized two-point, Schwenke-style~\cite{Schwenke05} formula,
\begin{equation}\label{eq:cbs_extrap}
E^{C}_{\mathrm{CBS}} = (E_{n+1} - E_{n})F^{C}_{n+1} + E_{n}
\end{equation}

\noindent originally proposed by Hill et al.~\cite{Hill09} The coefficients $F^{C}_{n+1}$ are specific to the $\mathrm{CCSD-F12b}$ and $\mathrm{(T)}$ components of the total CCSD(T)-F12b energy and we use values of $F^{\mathrm{CCSD-F12b}}=1.363388$ and $F^{\mathrm{(T)}}=1.769474$ as recommended in Ref.~\onlinecite{Hill09}. No extrapolation was applied to the Hartree-Fock (HF) energy, rather the HF+CABS singles correction~\cite{Adler07} calculated in the larger basis set was used.

 The energy correction from core-valence electron correlation $\Delta E_{\mathrm{CV}}$ was calculated at the CCSD(T)-F12b level of theory with the F12-optimized correlation consistent core-valence basis set cc-pCVQZ-F12.~\cite{Hill10} The same ansatz and ABS as in the frozen core approximation computations were used, however we set $\beta=1.5$~$a_0^{-1}$. All-electron calculations kept the (1\textit{s}) orbital of Cl frozen with all other electrons correlated due to the inability of the basis set to adequately describe this orbital.
 
 Core-valence and higher-order electron correlation often contribute to the electronic energy with opposing signs and should thus be considered jointly. We use the hierarchy of coupled cluster methods to estimate the HO correction as $\Delta E_{\mathrm{HO}} = \Delta E_{\mathrm{T}} + \Delta E_{\mathrm{(Q)}}$, including the full triples contribution $\Delta E_{\mathrm{T}} = \left[E_{\mathrm{CCSDT}}-E_{\mathrm{CCSD(T)}}\right]$, and the perturbative quadruples contribution $\Delta E_{\mathrm{(Q)}} = \left[E_{\mathrm{CCSDT(Q)}}-E_{\mathrm{CCSDT}}\right]$. Calculations were carried out in the frozen core approximation at the CCSD(T), CCSDT and CCSDT(Q) levels of theory using the general coupled cluster approach~\cite{Kallay05,Kallay08} as implemented in the MRCC code~\cite{mrcc} interfaced to CFOUR.~\cite{cfour} For the full triples and the perturbative quadruples calculations, we employed the augmented correlation consistent triple zeta basis set, aug-cc-pVTZ(+d for Cl),~\cite{Dunning89,Kendall92,Woon93,Dunning01} and the double zeta basis set, aug-cc-pVDZ(+d for Cl), respectively. Note that for HO coupled cluster corrections, it is possible to use successively smaller basis sets at each step up in excitation level due to faster convergence.~\cite{Feller06}
 
 In exploratory calculations, the contributions from the full quadruples $\left[E_{\mathrm{CCSDTQ}}-E_{\mathrm{CCSDT(Q)}}\right]$ and from the perturbative pentuples $\left[E_{\mathrm{CCSDTQ(P)}}-E_{\mathrm{CCSDTQ}}\right]$ were found to largely cancel each other out. Thus to reduce the computational expense, only $\Delta E_{\mathrm{T}}$ and $\Delta E_{\mathrm{(Q)}}$ were deemed necessary for an adequate representation of HO electron correlation.
 
 Scalar relativistic effects $\Delta E_{\mathrm{SR}}$ were included through the one-electron mass velocity and Darwin terms (MVD1) from the Breit-Pauli Hamiltonian in first-order perturbation theory.~\cite{Cowan76} Calculations were performed with all electrons correlated (except for the (1\textit{s}) of Cl) at the CCSD(T)/aug-cc-pCVTZ(+d for Cl)~\cite{Woon95,Peterson02} level of theory using the MVD1 approach~\cite{Klopper97} implemented in CFOUR. The contribution from the two-electron Darwin term is expected to be small enough to be neglected.~\cite{Gauss07}
 
 The diagonal Born-Oppenheimer correction $\Delta E_{\mathrm{DBOC}}$ was computed again with the (1\textit{s}) orbital of Cl frozen and all other electrons correlated. Calculations employed the CCSD method~\cite{Gauss06} as implemented in CFOUR with the aug-cc-pCVTZ(+d for Cl) basis set. The DBOC is the contribution from the nuclear kinetic energy operator acting on the ground state electronic wavefunction. It is mass dependent, so separate contributions were generated for CH$_{3}{}^{35}$Cl and CH$_{3}{}^{37}$Cl.

 The spin-orbit interaction was not considered as it can be safely neglected in spectroscopic calculations on light closed-shell molecules.~\cite{Tarczay01} A simple estimate of the Lamb shift was also calculated from the MVD1 contribution,~\cite{Pyykko01} but its effect on the vibrational energy levels was negligible. The differing levels of theory and basis set size reflect the fact that different HL energy corrections converge at different rates.

\begin{figure}
 \includegraphics{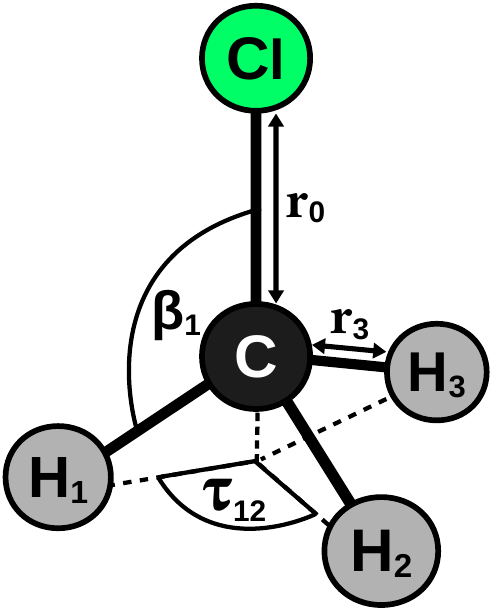}
 \caption{\label{fig:geometry}Definition of internal coordinates used for CH$_3$Cl.}
 \end{figure}

 Grid points were generated using a random energy-weighted sampling algorithm of Monte Carlo type, in terms of nine internal coordinates: the C-Cl bond length $r_0$; three C-H bond lengths $r_1$, $r_2$ and $r_3$; three $\angle(\mathrm{H}_i\mathrm{CCl})$ interbond angles $\beta_1$, $\beta_2$ and $\beta_3$; and two dihedral angles $\tau_{12}$ and $\tau_{13}$ between adjacent planes containing H$_i$CCl and H$_j$CCl (see Figure~\eqref{fig:geometry}). This led to a global grid of $44{\,}820$ geometries with energies up to $h c \cdot 50{\,}000{\,}$cm$^{-1}$, which included geometries in the range $1.3\leq r_0 \leq 2.95{\,}\mathrm{\AA}$, $0.7\leq r_i \leq 2.45{\,}\mathrm{\AA}$, $65\leq \beta_i \leq 165^{\circ}$ for $i=1,2,3$ and $55\leq \tau_{jk} \leq 185^{\circ}$ with $jk=12,13$. To ensure an adequate description of the equilibrium region, around $1000$ carefully chosen low-energy points were also incorporated into the data set. At each grid point, the computed coupled cluster energies were extrapolated to the CBS limit using Eq.\eqref{eq:cbs_extrap}.

\begin{figure*}
 \includegraphics{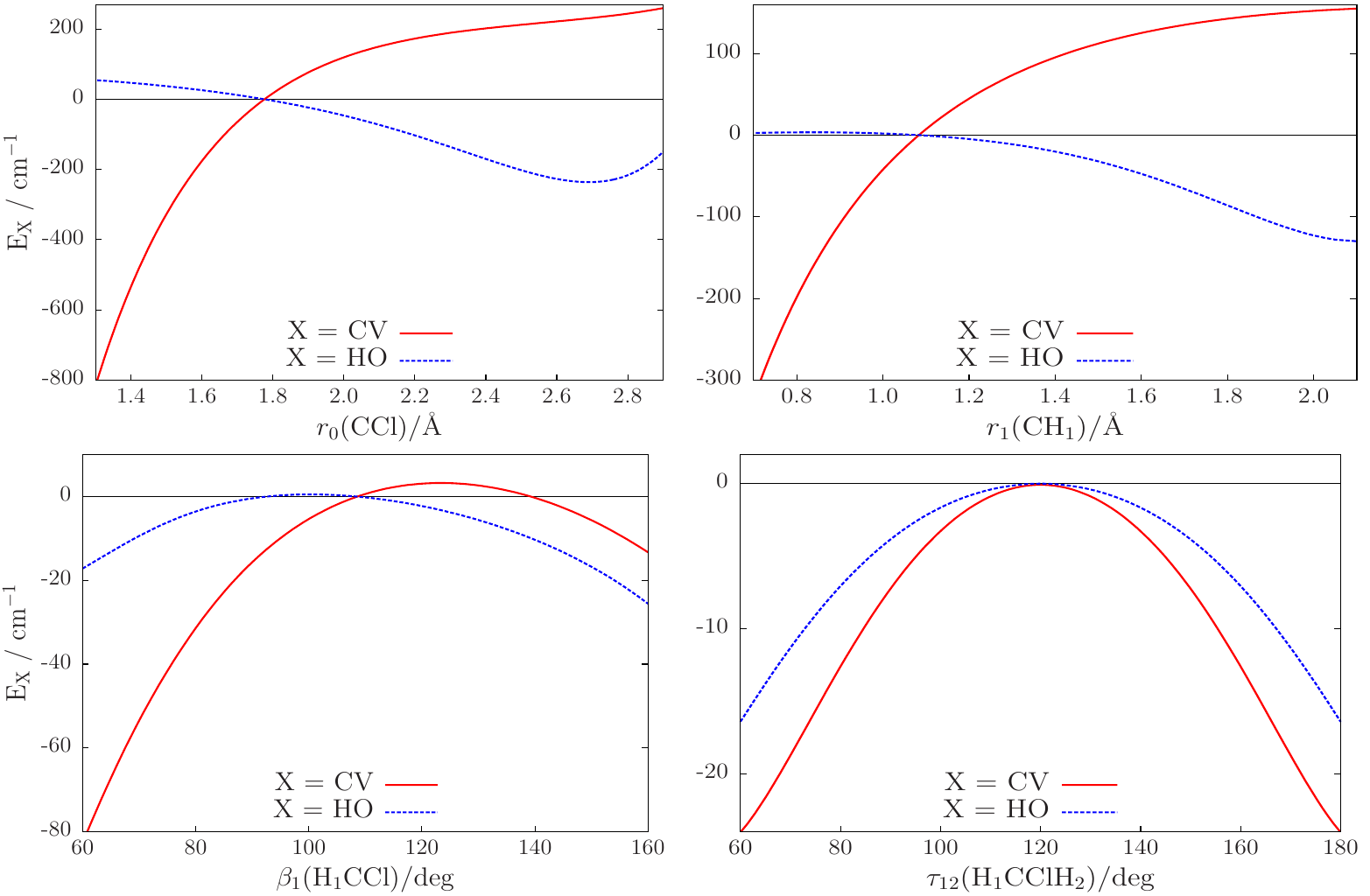}
 \caption{\label{fig:1d_cv_ho}One-dimensional cuts of the core-valence (CV) and higher-order (HO) corrections with all other coordinates held at their equilibrium values.}
 \end{figure*}

\begin{figure*}
 \includegraphics{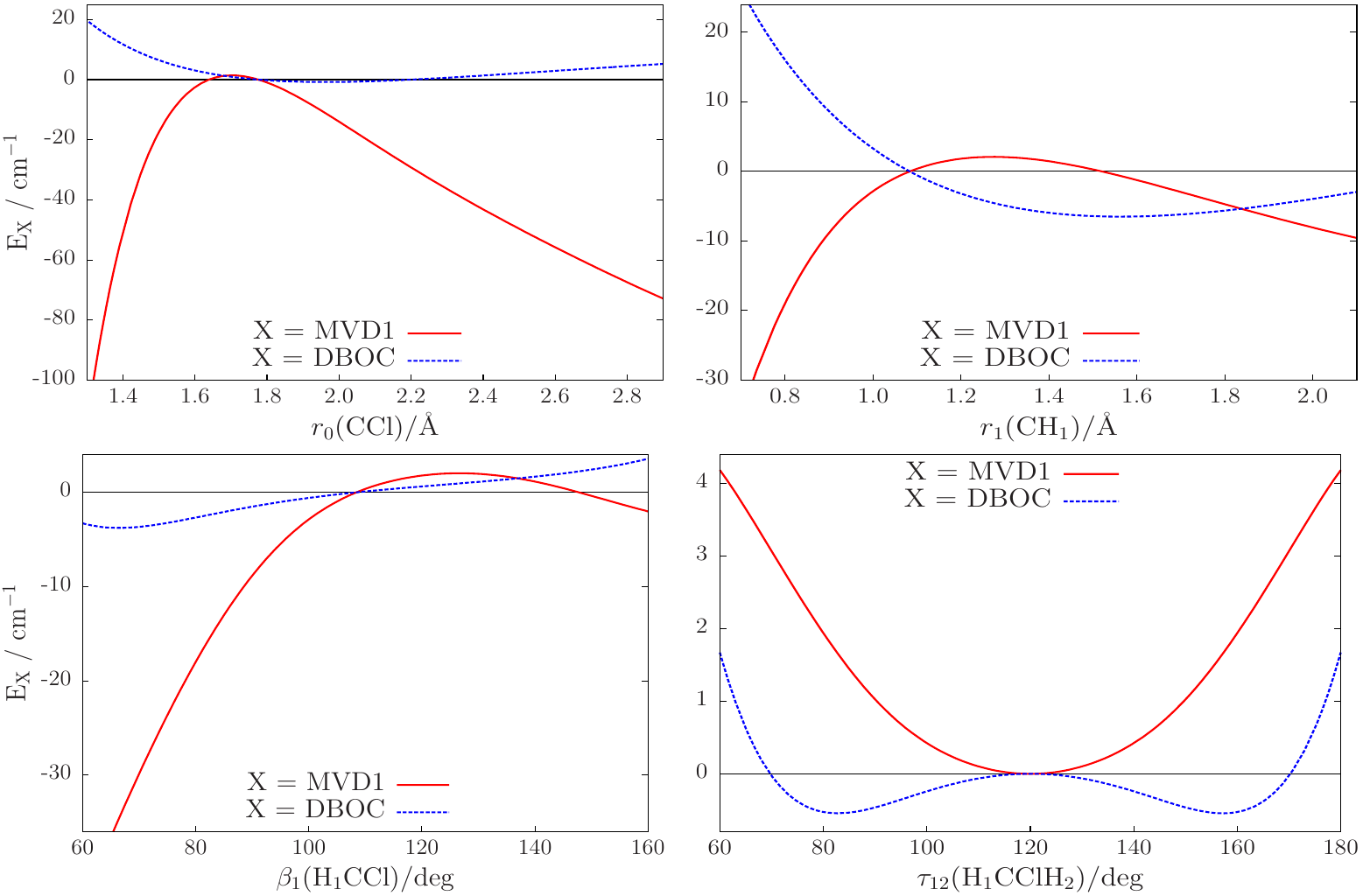}
 \caption{\label{fig:1d_mvd1_dboc} One-dimensional cuts of the scalar relativistic (MVD1) and diagonal Born-Oppenheimer (DBOC) corrections with all other coordinates held at their equilibrium values.}
 \end{figure*}

 The HL energy corrections are generally small in magnitude and vary in a smooth manner,~\cite{YaYuRi11.H2CS} displaying a straightforward polynomial-type dependence as can be seen in Figures \eqref{fig:1d_cv_ho} and \eqref{fig:1d_mvd1_dboc}. For each of the HL terms, a reduced grid was carefully designed to obtain a satisfactory description of the correction with minimum computational effort. Reduced grids of $9377$, $3526$, $12{\,}296$ and $3679$ points with energies up to $h c \cdot 50{\,}000{\,}$cm$^{-1}$ were used for the CV, HO, SR and DBOC corrections, respectively.
 
\subsection{Analytic representation}

 Methyl chloride is a prolate symmetric top molecule of the $\bm{C}_{3\mathrm{v}}\mathrm{(M)}$ symmetry group.~\cite{MolSym_BuJe98} Of the six symmetry operations $\left\lbrace E,(123),(132),(12)^{*},(23)^{*},(13)^{*}\right\rbrace$ which make up $\bm{C}_{3\mathrm{v}}$(M), the cyclic permutation $(123)$ replaces nucleus 1 with nucleus 2, nucleus 2 with nucleus 3, and nucleus 3 with nucleus 1. The permutation-inversion operation $(12)^{*}$ interchanges nuclei 1 and 2 and inverts all particles (including electrons) in the molecular centre of mass. The identity operation $E$ leaves the molecule unchanged.
 
 To represent the PES analytically, an on-the-fly symmetrization procedure has been implemented. We first introduce the coordinates
\begin{equation}\label{eq:stretch1}
\xi_1=1-\exp\left(-a(r_0 - r_0^{\mathrm{eq}})\right)
\end{equation}
\begin{equation}\label{eq:stretch2}
\xi_j=1-\exp\left(-b(r_i - r_1^{\mathrm{eq}})\right){\,};\hspace{2mm}j=2,3,4{\,}, \hspace{2mm} i=j-1
\end{equation}

\noindent where $a=1.65{\,}\mathrm{\AA}^{-1}$ for the C-Cl internal coordinate $r_0$, and $b=1.75{\,}\mathrm{\AA}^{-1}$ for the three C-H internal coordinates $r_1,r_2$ and $r_3$. For the angular terms
\begin{equation}\label{eq:angular1}
\xi_k = (\beta_i - \beta^{\mathrm{eq}}){\,};\hspace{2mm}k=5,6,7{\,}, \hspace{2mm} i=k-4
\end{equation}
\begin{equation}\label{eq:angular2}
\xi_8 = \frac{1}{\sqrt{6}}\left(2\tau_{23}-\tau_{13}-\tau_{12}\right)
\end{equation}
\begin{equation}\label{eq:angular3}
\xi_9 = \frac{1}{\sqrt{2}}\left(\tau_{13}-\tau_{12}\right)
\end{equation}

\noindent Here $\tau_{23}=2\pi-\tau_{12}-\tau_{13}$, and $r_0^{\mathrm{eq}}$, $r_1^{\mathrm{eq}}$ and $\beta^{\mathrm{eq}}$ are the reference equilibrium structural parameters of CH$_3$Cl. 

 Taking an initial potential term of the form
\begin{equation}\label{eq:V_i}
V_{ijk\ldots}^{\mathrm{initial}}=\xi_{1}^{\,i}\xi_{2}^{\,j}\xi_{3}^{\,k}\xi_{4}^{\,l}\xi_{5}^{\,m}\xi_{6}^{\,n}\xi_{7}^{\,p}\xi_{8}^{\,q}\xi_{9}^{\,r}
\end{equation}

\noindent with maximum expansion order $i+j+k+l+m+n+p+q+r=6$, each symmetry operation of $\bm{C}_{3\mathrm{v}}$(M) is independently applied to $V_{ijk\ldots}^{\mathrm{initial}}$, i.e.
\begin{equation}\label{eq:V_op}
V_{ijk\ldots}^{\mathbf{X}}=\mathbf{X}{\,}V_{ijk\ldots}^{\mathrm{initial}}=\mathbf{X}\left(\xi_{1}^{\,i}\xi_{2}^{\,j}\xi_{3}^{\,k}\xi_{4}^{\,l}\xi_{5}^{\,m}\xi_{6}^{\,n}\xi_{7}^{\,p}\xi_{8}^{\,q}\xi_{9}^{\,r}\right)
\end{equation}

\noindent where $\mathbf{X}=\lbrace E,(123),(132),(12)^{*},(23)^{*},(13)^{*}\rbrace$, to create six new terms. The results are summed up to produce a final term,
\begin{equation}\label{eq:V_f}
V_{ijk\ldots}^{\mathrm{final}}=V_{ijk\ldots}^{E}+V_{ijk\ldots}^{(123)}+V_{ijk\ldots}^{(132)}+V_{ijk\ldots}^{(12)^*}+V_{ijk\ldots}^{(23)^*}+V_{ijk\ldots}^{(13)^*}
\end{equation}

\noindent which is itself subjected to the six $\bm{C}_{3\mathrm{v}}$(M) symmetry operations to check its invariance. The total potential function is then given by the expression
\begin{equation}\label{eq:pot_f}
V_{\mathrm{total}}(\xi_{1},\xi_{2},\xi_{3},\xi_{4},\xi_{5},\xi_{6},\xi_{7},\xi_{8},\xi_{9})={\sum_{ijk\ldots}}{\,}\mathrm{f}_{ijk\ldots}V_{ijk\ldots}^{\mathrm{final}}
\end{equation}

\noindent where $\mathrm{f}_{ijk\ldots}$ are the corresponding expansion coefficients, determined through a least-squares fitting to the \textit{ab initio} data. Weight factors of the form suggested by Partridge and Schwenke,~\cite{Schwenke97}
\begin{equation}\label{eq:weights}
w_i=\left(\frac{\tanh\left[-0.0006\times(\tilde{E}_i - 15{\,}000)\right]+1.002002002}{2.002002002}\right)\times\frac{1}{N\tilde{E}_i^{(w)}}
\end{equation}

\noindent were used in the fitting, with normalization constant $N=0.0001$ and $\tilde{E}_i^{(w)}=\max(\tilde{E}_i, 10{\,}000)$, where $\tilde{E}_i$ is the potential energy at the $i$th geometry above equilibrium (all values in cm$^{-1}$). In our fitting, energies below $15{\,}000{\,}$cm$^{-1}$ are favoured by the weight factors. For geometries where $r_0\geq 2.35{\,}\mathrm{\AA}$ and $r_i\geq 2.00{\,}\mathrm{\AA}$ for $i=1,2,3$, the weights were reduced by several orders of magnitude. At such large stretch coordinates, the coupled cluster method is known to become unreliable, as indicated by a T1 diagnostic value $>0.02$.~\cite{T1_Lee89} Although energies at these points may not be wholly accurate, they are still useful and ensure that the PES maintains a reasonable shape towards dissociation.

 The same form of potential function, Eq.\eqref{eq:pot_f}, and the same procedure, Eqs.\eqref{eq:V_i} to \eqref{eq:V_f}, were used to fit the higher-level correction surfaces. The stretching coordinates however were replaced with linear expansion variables; $\xi_1=(r_0-r_0^{\mathrm{eq}})$ and $\xi_j=(r_i-r_1^{\mathrm{eq}})$ where $j=2,3,4$ and $i=j-1$. The angular terms, Eqs.\eqref{eq:angular1} to \eqref{eq:angular3}, remained the same as before. Each HL correction was fitted independently and the parameters $r_0^{\mathrm{eq}}$, $r_1^{\mathrm{eq}}$ and $\beta^{\mathrm{eq}}$ were optimized for each surface. The four HL corrections were applied at each of the $44{\,}820$ grid points, either from a directly calculated value at that geometry, or by interpolation using the corresponding analytic representation. Two final data sets were produced, one for each isotopologue of CH$_3$Cl, the only difference being the contribution from the DBOC.
 
 Two separate fits were carried out and in each instance we could usefully vary 414 expansion parameters to give a weighted root-mean-square (rms) error of $0.82{\,}$cm$^{-1}$ for energies up to $50{\,}000{\,}$cm$^{-1}$. The fit employed Watson's robust fitting scheme,~\cite{Watson03} the idea of which is to reduce the weight of outliers and lessen their influence in determining the final set of parameters. The Watson scheme improves the fit at energies below $10{\,}000{\,}$cm$^{-1}$ which is preferable for our purposes. When comparing the expansion parameters for CH$_{3}{}^{35}$Cl and CH$_{3}{}^{37}$Cl, only very slight differences were observed in the determined values. We refer to these two PESs as CBS-35$^{\,\mathrm{HL}}$ and CBS-37$^{\,\mathrm{HL}}$ in subsequent calculations.
    
 To assess the combined effect of the HL corrections and CBS extrapolation on the vibrational energy levels and equilibrium geometry of CH$_3$Cl, we fit a reference PES to the raw CCSD(T)-F12b/cc-pVQZ-F12 energies. Again we used Watson's robust fitting scheme and 414 parameters to give a weighted rms error of $0.82{\,}$cm$^{-1}$ for energies up to $50{\,}000{\,}$cm$^{-1}$. We refer to this PES as VQZ-F12 in subsequent calculations. Note that the CBS-(35/37)$^{\,\mathrm{HL}}$ and VQZ-F12 PESs contain only slightly different parameter sets.
 
 The choice of reference equilibrium structural parameters in our PES expansion is to some extent arbitrary due to the inclusion of linear expansion terms in the parameter set. For this reason, values of $r_0^{\mathrm{eq}}=1.7775{\,}\mathrm{\AA}$, $r_1^{\mathrm{eq}}=1.0837{\,}\mathrm{\AA}$, and $\beta^{\mathrm{eq}}=108.445^{\circ}$, used for the CBS-35$^{\,\mathrm{HL}}$ PES, were also employed for the CBS-37$^{\,\mathrm{HL}}$ and VQZ-F12 PESs. Note that these are not the actual equilibrium parameters which define the minimum of the PES, they are simply parameters of a function. The true equilibrium values will be determined and discussed in Sec.~\ref{sec:variational}.
 
 Generating a PES on-the-fly is advantageous when it comes to variational calculations as its implementation requires only a short amount of code. Alternatively, one can derive the full analytic expression for the potential and incorporate this into the nuclear motion computations, but this method is cumbersome. The CBS-35$^{\,\mathrm{HL}}$ and CBS-37$^{\,\mathrm{HL}}$ expansion parameter sets are provided in the supplementary material along with a FORTRAN routine to construct the PESs.~\cite{EPAPSCH3CL} 

\section{Results}
\label{sec:variational}

\subsection{Extrapolation to the Complete Vibrational Basis Set Limit}
\label{sec:cvbs}

 The nuclear motion code TROVE (Theoretical ROVibrational Energies)~\cite{TROVE2007} is designed to calculate the rotation-vibration energy levels and corresponding transition intensities for a polyatomic molecule of arbitrary structure in an isolated electronic state. The flexibility of TROVE has allowed a range of molecular systems to be treated,~\cite{OvThYu08a.PH3,OvThYu08.PH3,YaYuRi11.H2CS,YuBaYa09.NH3,YuYaTh09.HSOH,YaYuJe10.HSOH,YuCaYa10.SBH3,
 YaYuJe11.H2CO,UnTeYu13.SO3,PoKoOv13.H2O2,YuTeBa13.CH4} and for the present study the functionality has been extended to handle molecules of the form XY$_3$Z.

 In TROVE, solution of the rotation-vibration Schr\"{o}dinger equation is achieved by numerical diagonalization of the corresponding Hamiltonian constructed in terms of a symmetry adapted basis set. The rovibrational Hamiltonian is represented as a power series expansion around a reference geometry, taken presently at the equilibrium configuration. For the present work we take advantage of recent developments in TROVE, in particular the implementation of a novel method of constructing the rovibrational Hamiltonian in terms of curvilinear internal coordinates.~\cite{YaYu15.ADF} By employing this new approach, our variational results show much faster convergence with respect to vibrational basis set size. We will see the importance of this later on. For CH$_3$Cl, we truncate the kinetic and potential energy operators at 6th and 8th order respectively in all calculations. This level of truncation is sufficient for our purposes, however we refer the reader to Ref.~\onlinecite{TROVE2007} for a detailed discussion of the associated errors of such a scheme. Note that atomic mass values have been used in the subsequent TROVE computations.
 
 A multi-step contraction scheme is used to generate the vibrational basis set, the size of which is controlled by the polyad number 
\begin{equation}
 P = \sum_{k=1}^{9} a_k n_k
\end{equation}

\noindent The quantum numbers $n_k$ correspond to primitive basis functions $\phi_{n_k}$, which are obtained from solving one-dimensional Schr\"{o}dinger equations for each of the nine vibrational modes by means of the Numerov-Cooley method.~\cite{Numerov1924,Cooley1961} Using the definition of the polyad coefficient $a_k=\omega_k/\min(\omega_1\ldots\omega_9)$, where $\omega_k$ denotes the harmonic frequency of the $k$th mode, the polyad number for CH$_{3}$Cl is
\begin{equation}\label{eq:polyad}
P = n_1+2(n_2+n_3+n_4)+n_5+n_6+n_7+n_8+n_9 \leq P_{\mathrm{max}}
\end{equation}

\noindent which does not exceed a predefined maximum value $P_{\mathrm{max}}$.

\begin{figure}
 \includegraphics{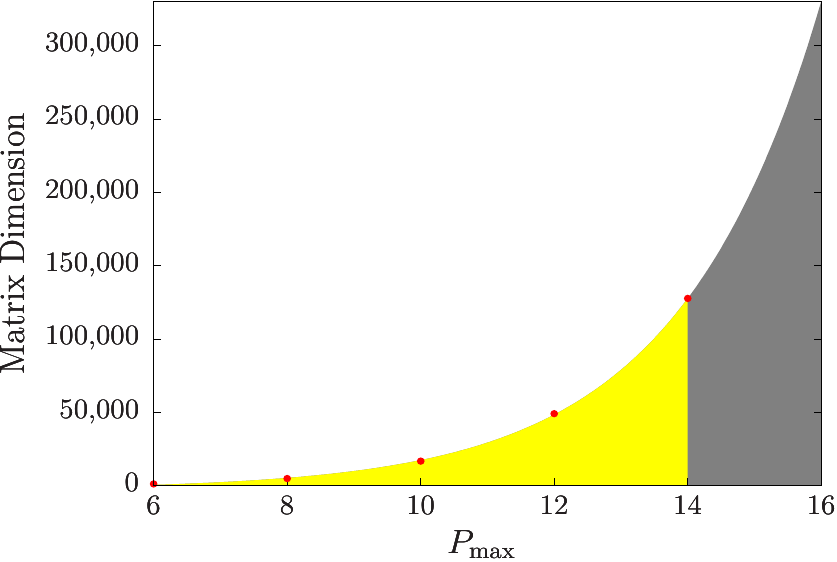}
 \caption{\label{fig:dimension}Size of the $J=0$ Hamiltonian matrix with respect to the polyad truncation number $P_{\mathrm{max}}$. Computations were feasible up to $P_{\mathrm{max}}=14$.}
 \end{figure}

 Fully converged energies in variational calculations are usually obtained with the use of an extended basis set. We have only been able to compute $J=0$ vibrational energies up to a polyad truncation number of $P_{\mathrm{max}}=14$ for CH$_3$Cl. As shown in Figure~\eqref{fig:dimension}, this requires the diagonalization of a Hamiltonian matrix of dimension close to $128{\,}000$, which in turn equals the number of primitive basis functions generated. The extension to $P_{\mathrm{max}}=16$ using TROVE would be an arduous computational task. 
 
 One means of achieving converged vibrational energy levels without having to diagonalize increasingly large matrices, is the use of a complete vibrational basis set (CVBS) extrapolation.~\cite{OvThYu08.PH3} In analogy to the common basis set extrapolation techniques of electronic structure theory,~\cite{Petersson88,Petersson91} the same principles can be applied to TROVE calculations with respect to $P_{\mathrm{max}}$. We adopt the exponential decay expression
\begin{equation}
 E_i(P_{\mathrm{max}}) = E_i^{\mathrm{CVBS}}+A_i\exp(-\lambda_i P_{\mathrm{max}})
\end{equation}

\noindent where $E_i$ is the energy of the $i$th level, $E_i^{\mathrm{CVBS}}$ is the respective energy at the CVBS limit, $A_i$ is a fitting parameter, and $\lambda_i$ can be found from
\begin{equation}
\lambda_i=-\frac{1}{2}\ln\left(\frac{E_i(P_{\mathrm{max}}+2)-E_i(P_{\mathrm{max}})}{E_i(P_{\mathrm{max}})-E_i(P_{\mathrm{max}}-2)}\right)
\end{equation}

\begin{figure}
 \includegraphics{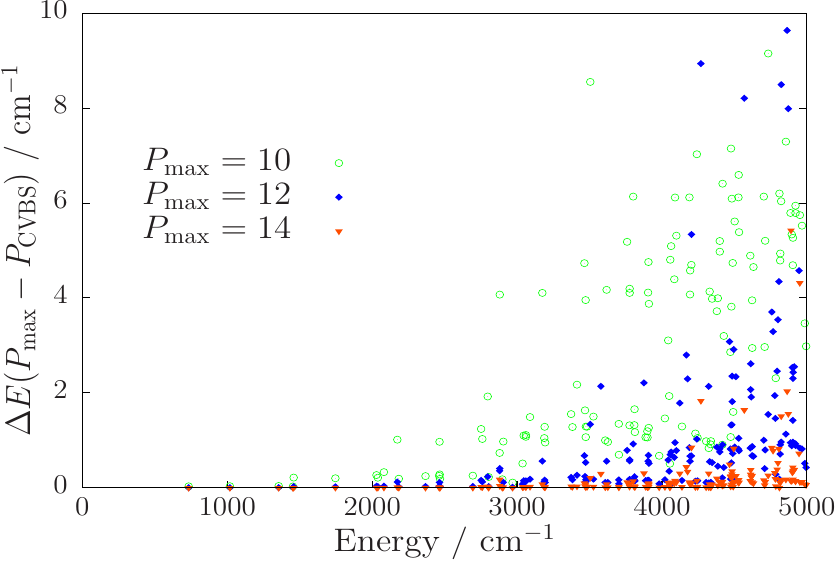}
 \caption{\label{fig:converge}Convergence of vibrational term values of CH$_{3}{}^{35}$Cl up to $5000{\,}$cm$^{-1}$ with respect to $P_{\mathrm{max}}=P_{\mathrm{CVBS}}$. For illustrative purposes we restrict the range of $\Delta E(P_{\mathrm{max}}-P_{\mathrm{CVBS}})$ to $10{\,}$cm$^{-1}$.}
 \end{figure}

 Values of $P_{\mathrm{max}}=\lbrace 10,12,14 \rbrace$ were employed for a CVBS extrapolation of all vibrational term values up to $5000{\,}$cm$^{-1}$, and for selected higher energies to compare with experiment. This was done for the CBS-35$^{{\,}\mathrm{HL}}$, CBS-37$^{{\,}\mathrm{HL}}$, and VQZ-F12 PESs. In Figure~\eqref{fig:converge}, the convergence of the vibrational energy levels up to $5000{\,}$cm$^{-1}$ for the CBS-35$^{{\,}\mathrm{HL}}$ PES can be seen with respect to the final $E_i^{\mathrm{CVBS}}$ extrapolated values. Below $4000{\,}$cm$^{-1}$ the computed $P_{\mathrm{max}}=14$ term values are already reasonably well converged. Only five levels in this range possess a residual $\Delta E(P_{\mathrm{max}}-P_{\mathrm{CVBS}})$ larger than $0.1{\,}$cm$^{-1}$, none of which is greater than $0.3{\,}$cm$^{-1}$. As expected, levels involving highly excited modes benefit the most from extrapolation as these converge at a much slower rate. 

 The limiting factor of a CVBS extrapolation is the correct identification of the energy levels at each step up in basis set size. TROVE automatically assigns quantum numbers to the eigenvalues and corresponding eigenvectors by analysing the contribution of the basis functions. Due to the increased density of states above $5000{\,}$cm$^{-1}$ for higher values of $P_{\mathrm{max}}$, it quickly becomes difficult to consistently identify and match levels, except for highly excited individual modes. 

\subsection{Vibrational $J=0$ Energies}
\label{sec:vib_enr}

 The normal modes of methyl chloride are classified by two symmetry species, $A_{1}$ and $E$. Of $A_{1}$ symmetry are the three non-degenerate modes; the symmetric CH$_{3}$ stretching mode $\nu_{1}$ ($2967.77/2967.75{\,}$cm$^{-1}$), the symmetric CH$_{3}$ deformation mode $\nu_{2}$ ($1354.88/1354.69{\,}$cm$^{-1}$), and the C{--}Cl stretching mode $\nu_{3}$ ($732.84/727.03{\,}$cm$^{-1}$). Of $E$ symmetry are the three degenerate modes; the CH$_{3}$ stretching mode $\nu_{4}^{l_{4}}$ ($3039.26/3039.63{\,}$cm$^{-1}$), the CH$_{3}$ deformation mode $\nu_{5}^{l_{5}}$ ($1452.18/1452.16{\,}$cm$^{-1}$), and the CH$_{3}$ rocking mode $\nu_{6}^{l_{6}}$ ($1018.07/1017.68{\,}$cm$^{-1}$). The values in parentheses are the experimentally determined values for CH$_{3}{}^{35}$Cl/CH$_{3}{}^{37}$Cl from Refs.~\onlinecite{11BrPeJa.CH3Cl} and \onlinecite{05NiChBu.CH3Cl}. The additional vibrational angular momentum quantum numbers $l_{4}$, $l_{5}$, and $l_{6}$ are needed to resolve the degeneracy of their respective modes. To be of spectroscopic use we map the vibrational quantum numbers $n_k$ of TROVE to the normal mode quantum numbers $\mathrm{v}_k$ commonly used. For CH$_3$Cl, the vibrational states are labelled as $\mathrm{v_1}\nu_1+\mathrm{v_2}\nu_2+\mathrm{v_3}\nu_3+\mathrm{v_4}\nu_4+\mathrm{v_5}\nu_5+\mathrm{v_6}\nu_6$ where $\mathrm{v_i}$ counts the level of excitation.

 The calculated $J=0$ energy levels for CH$_{3}{}^{35}$Cl using the CBS-35$^{{\,}\mathrm{HL}}$ and VQZ-F12 PESs are listed in Table~\ref{tab:j0_cbs_35cl}. We compare against all available experimental data taken from Refs.~\onlinecite{11BrPeJa.CH3Cl}, \onlinecite{05NiChBu.CH3Cl}, \onlinecite{90DuLaxx.CH3Cl}, and \onlinecite{99Laxxxx.CH3Cl}. A small number of levels from Refs.~\onlinecite{90DuLaxx.CH3Cl} and \onlinecite{99Laxxxx.CH3Cl} have not been included as we were unable to confidently identify the corresponding values in TROVE.
 
\setlength\LTleft{0pt}
\setlength\LTright{0pt}
\LTcapwidth=\textwidth
\begin{longtable*}[ht]{@{\extracolsep{\fill}} l c c c c c c c}
\caption{\label{tab:j0_cbs_35cl} Comparison of calculated and experimental $J=0$ vibrational term values (in cm$^{-1}$) for CH$_{3}{}^{35}$Cl. The zero-point energy was computed to be $8219.661\,$cm$^{-1}$ at the CVBS limit.}\\ \hline\hline
Mode & Sym. & VQZ-F12{\,}(A) & CBS-35$^{{\,}\mathrm{HL}}${\,}(B) & Exp. & Obs$-$calc{\,}(A) & Obs$-$calc{\,}(B) & Ref.\\ \hline
\endfirsthead
\caption{(\textit{Continued})}\\ \hline 
Mode & Sym. & VQZ-F12{\,}(A) & CBS-35$^{{\,}\mathrm{HL}}${\,}(B) & Exp. & Obs$-$calc{\,}(A) & Obs$-$calc{\,}(B) & Ref.\\ \hline
\endhead
$\nu_3$       & $A_1$ & 734.37  &  733.22 & 732.8422  & -1.53 & -0.38 & \onlinecite{05NiChBu.CH3Cl}\\
$\nu_6$       & $E$   & 1018.16 & 1018.05 & 1018.0709 & -0.09 &  0.02 & \onlinecite{05NiChBu.CH3Cl}\\
$\nu_2$       & $A_1$ & 1355.11 & 1355.01 & 1354.8811 & -0.23 & -0.13 & \onlinecite{05NiChBu.CH3Cl}\\
$\nu_5$       & $E$   & 1451.57 & 1452.56 & 1452.1784 &  0.61 & -0.38 & \onlinecite{05NiChBu.CH3Cl}\\
$2\nu_3$      & $A_1$ & 1459.94 & 1457.54 & 1456.7626 & -3.18 & -0.78 & \onlinecite{05NiChBu.CH3Cl}\\
$\nu_3+\nu_6$ & $E$   & 1747.10 & 1745.78 & 1745.3711 & -1.73 & -0.41 & \onlinecite{05NiChBu.CH3Cl}\\
$2\nu_6$      & $A_1$ & 2029.67 & 2029.46 & 2029.3753 & -0.29 & -0.09 & \onlinecite{05NiChBu.CH3Cl}\\
$2\nu_6$      & $E$   & 2038.58 & 2038.37 & 2038.3262 & -0.25 & -0.04 & \onlinecite{05NiChBu.CH3Cl}\\
$\nu_2+\nu_3$ & $A_1$ & 2082.35 & 2080.98 & 2080.5357 & -1.82 & -0.45 & \onlinecite{05NiChBu.CH3Cl}\\
$3\nu_3$      & $A_1$ & 2176.84 & 2173.09 & 2171.8875 & -4.95 & -1.20 & \onlinecite{05NiChBu.CH3Cl}\\
$\nu_3+\nu_5$ & $E$   & 2183.51 & 2183.30 & 2182.5717 & -0.94 & -0.73 & \onlinecite{05NiChBu.CH3Cl}\\
$\nu_2+\nu_6$ & $E$   & 2368.08 & 2367.90 & 2367.7222 & -0.35 & -0.18 & \onlinecite{05NiChBu.CH3Cl}\\
$\nu_5+\nu_6$ & $E$   & 2461.19 & 2461.98 & 2461.6482 &  0.46 & -0.33 & \onlinecite{05NiChBu.CH3Cl}\\
$2\nu_3+\nu_6$& $E$   & 2467.19 & 2464.65 & 2463.8182 & -3.37 & -0.83 & \onlinecite{05NiChBu.CH3Cl}\\
$\nu_5+\nu_6$ & $A_1$ & 2464.50 & 2465.28 & 2464.9025 &  0.40 & -0.38 & \onlinecite{05NiChBu.CH3Cl}\\
$\nu_5+\nu_6$ & $A_2$ & 2466.85 & 2467.85 & 2467.6694 &  0.82 & -0.18 & \onlinecite{05NiChBu.CH3Cl}\\
$2\nu_2$      & $A_1$ & 2694.69 & 2694.61 & 2693.00   & -1.69 & -1.61 & \onlinecite{90DuLaxx.CH3Cl}\\
$\nu_3+2\nu_6$& $A_1$ & 2753.23 & 2751.74 & 2751.18   & -2.05 & -0.56 & \onlinecite{90DuLaxx.CH3Cl}\\
$\nu_2+2\nu_3$& $A_1$ & 2800.39 & 2797.64 & 2796.81   & -3.58 & -0.83 & \onlinecite{90DuLaxx.CH3Cl}\\
$\nu_2+\nu_5$ & $E$   & 2803.10 & 2803.96 & 2803.26   &  0.16 & -0.70 & \onlinecite{90DuLaxx.CH3Cl}\\
$4\nu_3$      & $A_1$ & 2885.23 & 2880.47 & 2878.00   & -7.23 & -2.47 & \onlinecite{90DuLaxx.CH3Cl}\\
$2\nu_5$      & $A_1$ & 2877.75 & 2879.31 & 2879.25   &  1.50 & -0.06 & \onlinecite{90DuLaxx.CH3Cl}\\
$2\nu_5$      & $E$   & 2896.27 & 2898.22 & 2895.566  & -0.71 & -2.65 & \onlinecite{11BrPeJa.CH3Cl}\\
$2\nu_3+\nu_5$& $E$   & 2906.64 & 2905.14 & 2907.903  &  1.26 &  2.77 & \onlinecite{11BrPeJa.CH3Cl}\\
$\nu_1$       & $A_1$ & 2965.78 & 2969.16 & 2967.7691 &  1.99 & -1.39 & \onlinecite{11BrPeJa.CH3Cl}\\
$\nu_4$       & $E$   & 3035.50 & 3038.19 & 3039.2635 &  3.76 &  1.07 & \onlinecite{11BrPeJa.CH3Cl}\\
$3\nu_6$      & $E$   & 3045.08 & 3045.76 & 3042.8944 & -2.18 & -2.87 & \onlinecite{11BrPeJa.CH3Cl}\\
$3\nu_6$      & $A_1$ & 3060.95 & 3060.62 & 3060.0064 & -0.95 & -0.62 & \onlinecite{11BrPeJa.CH3Cl}\\
$\nu_2+2\nu_6$& $A_1$ & 3373.81 & 3373.57 & 3373.5    & -0.31 & -0.07 & \onlinecite{90DuLaxx.CH3Cl}\\
$2\nu_2+\nu_3$& $A_1$ & 3415.53 & 3414.02 & 3413.0    & -2.53 & -1.02 & \onlinecite{90DuLaxx.CH3Cl}\\
$\nu_3+2\nu_5$& $A_1$ & 3607.98 & 3608.77 & 3607.70   & -0.28 & -1.07 & \onlinecite{90DuLaxx.CH3Cl}\\
$\nu_1+\nu_3$ & $A_1$ & 3700.18 & 3702.43 & 3700.67   &  0.49 & -1.76 & \onlinecite{90DuLaxx.CH3Cl}\\
$2\nu_2+\nu_6$& $E$   & 3702.92 & 3702.80 & 3702.69   & -0.23 & -0.11 & \onlinecite{90DuLaxx.CH3Cl}\\
$\nu_3+3\nu_6$& $E$   & 3760.53 & 3759.07 & 3756.6    & -3.93 & -2.47 & \onlinecite{90DuLaxx.CH3Cl}\\
$\nu_3+\nu_4$ & $E$   & 3773.98 & 3776.04 & 3773.52   & -0.46 & -2.52 & \onlinecite{90DuLaxx.CH3Cl}\\
$2\nu_5+\nu_6$& $E$   & 3884.88 & 3886.75 & 3886.05   &  1.17 & -0.70 & \onlinecite{90DuLaxx.CH3Cl}\\
$\nu_1+\nu_6$ & $E$   & 3977.68 & 3980.97 & 3979.66   &  1.98 & -1.31 & \onlinecite{90DuLaxx.CH3Cl}\\
$\nu_4+\nu_6$ & $E$   & 4047.37 & 4049.83 & 4051.22   &  3.85 &  1.39 & \onlinecite{90DuLaxx.CH3Cl}\\
$2\nu_2+\nu_5$& $E$   & 4137.96 & 4138.86 & 4138.29   &  0.33 & -0.57 & \onlinecite{90DuLaxx.CH3Cl}\\
$\nu_2+2\nu_5$& $A_1$ & 4229.37 & 4231.18 & 4230.34   &  0.97 & -0.84 & \onlinecite{90DuLaxx.CH3Cl}\\
$\nu_2+3\nu_6$& $E$  & 4378.39 & 4379.54 & 4380.52   &  2.13 &  0.98 & \onlinecite{90DuLaxx.CH3Cl}\\
$\nu_2+\nu_4$ & $E$   & 4384.03 & 4385.90 & 4382.64   & -1.39 & -3.26 & \onlinecite{90DuLaxx.CH3Cl}\\
$\nu_1+\nu_5$ & $E$   & 4412.81 & 4416.81 & 4415.4    &  2.59 & -1.41 & \onlinecite{99Laxxxx.CH3Cl}\\
$\nu_1+2\nu_6$& $A_1$ & 4982.68 & 4985.90 & 4984.0    &  1.32 & -1.90 & \onlinecite{90DuLaxx.CH3Cl}\\
$\nu_1+2\nu_2$& $A_1$ & 5655.67 & 5658.93 & 5657.0    &  1.33 & -1.93 & \onlinecite{90DuLaxx.CH3Cl}\\
$2\nu_2+\nu_4$& $E$   & 5708.85$^a$& 5711.12$^a$&5713 &  4.15 &  1.88 & \onlinecite{90DuLaxx.CH3Cl}\\
$\nu_1+\nu_4$ & $E$   & 5870.30 & 5875.98 & 5873.8    &  3.50 & -2.18 & \onlinecite{99Laxxxx.CH3Cl}\\
$2\nu_1$      & $A_1$ & 5875.28 & 5881.04 & 5878      &  2.72 & -3.04 & \onlinecite{99Laxxxx.CH3Cl}\\
$\nu_4+2\nu_5$& $E$   & 5918.20$^a$&5923.37$^a$&5923.4&  5.20 &  0.03 & \onlinecite{99Laxxxx.CH3Cl}\\
$2\nu_4$      & $A_1$ & 6011.38 & 6018.47 & 6015.3    &  3.92 & -3.17 & \onlinecite{90DuLaxx.CH3Cl}\\
$2\nu_1+\nu_5$& $E$   & 7303.96 & 7311.08 & 7313.2    &  9.24 &  2.12 & \onlinecite{90DuLaxx.CH3Cl}\\
$2\nu_4+\nu_5$& $E$   & 7437.80 & 7445.86 & 7443.2    &  5.40 & -2.66 & \onlinecite{90DuLaxx.CH3Cl}\\
$2\nu_4+2\nu_5$& $A_1$& 8870.15 & 8877.25 & 8874.3    &  4.15 & -2.95 & \onlinecite{90DuLaxx.CH3Cl}\\
$3\nu_4$      & $A_1$ & 9069.53 & 9079.28 & 9076.9    &  7.37 & -2.38 & \onlinecite{90DuLaxx.CH3Cl}\\
\hline\hline
$^a$ $P_{\mathrm{max}}=14$ value. \\[-2mm]
\end{longtable*}
 
 The CBS-35$^{{\,}\mathrm{HL}}$ PES reproduces the six fundamental term values with a root-mean-square (rms) error of $0.75{\,}$cm$^{-1}$ and a mean-absolute-deviation (mad) of $0.56{\,}$cm$^{-1}$. This is a considerable improvement over the results of the VQZ-F12 PES, which reproduces the fundamentals with a rms error of $1.86{\,}$cm$^{-1}$ and a mad of $1.37{\,}$cm$^{-1}$. Inspection of all computed CH$_{3}{}^{35}$Cl energy levels shows that on the whole, the CBS-35$^{{\,}\mathrm{HL}}$ PES is more reliable. This is gratifying as the effort required to generate the CBS-35$^{{\,}\mathrm{HL}}$ PES is far greater than that of the VQZ-F12 PES. Unlike other instances,~\cite{YaYuRi11.H2CS} the VQZ-F12 results do not benefit from an extensive cancellation of errors. Note that the PES reported in Ref.~\onlinecite{08Nixxxx.CH3Cl}, which did not treat any additional HL energy corrections, produces results with errors similar to those of the VQZ-F12 PES.
 
 The accuracy achieved at lower energies with the CBS-35$^{{\,}\mathrm{HL}}$ PES is quite remarkable, with residuals larger than $2{\,}$cm$^{-1}$ starting to appear around $3000{\,}$cm$^{-1}$. This is a notoriously difficult region of CH$_{3}$Cl with strong resonances, but the experimental values we compare against are from a recent high-resolution study and should thus be trustworthy.~\cite{11BrPeJa.CH3Cl} In the comparison against values reported in Ref.~\onlinecite{90DuLaxx.CH3Cl}, and subsequently used in Ref.~\onlinecite{99Laxxxx.CH3Cl}, we exercise some caution. The $2\nu_5(E)$ and $2\nu_3+\nu_5(E)$ levels presented in Ref.~\onlinecite{90DuLaxx.CH3Cl} are lower by around $3$ and $5{\,}$cm$^{-1}$ respectively when compared with new values measured in Ref.~\onlinecite{11BrPeJa.CH3Cl}. However the agreement for the $\nu_1$, $\nu_4$ and $3\nu_6(E)$ levels is excellent. The residual for the $2\nu_2$ level seems large given the residual for the $\nu_2$ term value, and we suspect that the experimental value is incorrect. At higher energies the quality of the CBS-35$^{{\,}\mathrm{HL}}$ PES does not appear to deteriorate significantly.
 
 For the $^{37}$Cl isotopologue of methyl chloride, the $J=0$ term values calculated from the CBS-37$^{{\,}\mathrm{HL}}$ PES are compared with all available experimental data in Table~\ref{tab:j0_cbs_37cl}. The CBS-37$^{{\,}\mathrm{HL}}$ PES reproduces the six fundamental term values with a rms error of $1.00{\,}$cm$^{-1}$ and a mad of $0.70{\,}$cm$^{-1}$. The reduction in accuracy when compared to the CBS-35$^{{\,}\mathrm{HL}}$ PES is primarily due to the $\nu_4$ mode, whose residual has gone from $1.07{\,}$cm$^{-1}$ for CH$_{3}{}^{35}$Cl to $1.92{\,}$cm$^{-1}$ for CH$_{3}{}^{37}$Cl. The accuracy of the $3\nu_6(E)$ level has also declined, but for energies leading up to $3000{\,}$cm$^{-1}$ the agreement with experiment is excellent. Despite being unable to compare against higher energies we expect the CBS-37$^{{\,}\mathrm{HL}}$ PES to perform as well as its $^{35}$Cl counterpart.
 
\begin{table}[!ht]
\tabcolsep=5pt
\caption{\label{tab:j0_cbs_37cl} Comparison of calculated and experimental $J=0$ vibrational term values (in cm$^{-1}$) for CH$_{3}{}^{37}$Cl. The zero-point energy was computed to be $8216.197\,$cm$^{-1}$ at the CVBS limit.}
\begin{center}
	\begin{tabular}{lcccc}
	\hline\hline
 	Mode &  Sym. & CBS-37$^{{\,}\mathrm{HL}}$ & Exp.$^a$ & Obs$-$calc \\
	\hline
	
	$\nu_3$       & $A_1$ &  727.40 &  727.0295 & -0.37 \\
	$\nu_6$       & $E$   & 1017.66 & 1017.6824 &  0.02 \\
	$\nu_2$       & $A_1$ & 1354.82 & 1354.6908 & -0.13 \\
	$2\nu_3$      & $A_1$ & 1446.12 & 1445.3509 & -0.77 \\
	$\nu_5$       & $E$   & 1452.53 & 1452.1552 & -0.38 \\
	$\nu_3+\nu_6$ & $E$   & 1739.64 & 1739.2357 & -0.41 \\
	$2\nu_6$      & $A_1$ & 2028.68 & 2028.5929 & -0.09 \\
	$2\nu_6$      & $E$   & 2037.59 & 2037.5552 & -0.04 \\
	$\nu_2+\nu_3$ & $A_1$ & 2074.90 & 2074.4526 & -0.45 \\
	$3\nu_3$      & $A_1$ & 2156.31 & 2155.1179 & -1.19 \\
	$\nu_3+\nu_5$ & $E$   & 2177.47 & 2176.7504 & -0.72 \\
	$\nu_2+\nu_6$ & $E$   & 2367.32 & 2367.1394 & -0.18 \\
	$2\nu_3+\nu_6$& $E$   & 2452.76 & 2451.9048 & -0.85 \\
	$\nu_5+\nu_6$ & $E$   & 2461.78 & 2461.4849 & -0.29 \\
	$\nu_5+\nu_6$ & $A_1$ & 2464.85 & 2464.4690 & -0.38 \\
	$\nu_5+\nu_6$ & $A_2$ & 2467.43 & 2467.2469 & -0.18 \\
	$\nu_2+\nu_5$ & $E$   & 2803.73 & 2803.2$\,^b$ & -0.53 \\
    $2\nu_5$      & $A_1$ & 2879.81 & 2879.0$\,^b$ & -0.81 \\
	$2\nu_3+\nu_5$& $E$   & 2893.71 & 2893.7394$\,^c$ & 0.03 \\
	$2\nu_5$      & $E$   & 2898.19 & 2895.449$\,^c$ & -2.74 \\
	$\nu_1$       & $A_1$ & 2969.14 & 2967.7469$\,^c$ & -1.39 \\
	$\nu_4$       & $E$   & 3037.71 & 3039.6311$\,^c$ &  1.92 \\
	$3\nu_6$      & $E$   & 3044.97 & 3041.2568$\,^c$ & -3.72 \\
	$3\nu_6$      & $A_1$ & 3059.47 & 3058.6913$\,^c$ & -0.78 \\
	\hline\hline
    \end{tabular}
\end{center}
 $^a$ Values from Ref.~\onlinecite{05NiChBu.CH3Cl} unless stated otherwise.\\[-2mm]
 $^b$ Ref.~\onlinecite{82BeAlGu.CH3Cl}.
 $^c$ Ref.~\onlinecite{11BrPeJa.CH3Cl}.\\[-2mm]
\end{table}
 
 We have not computed term values for CH$_{3}{}^{37}$Cl using the VQZ-F12 PES but we expect errors similar to those reported for CH$_{3}{}^{35}$Cl. It is evident that for methyl chloride, the inclusion of additional HL corrections and a CBS extrapolation in the PES lead to considerable improvements in computed $J=0$ energies. We have been able to identify and assign over $100$ new energy levels for both CH$_{3}{}^{35}$Cl and CH$_{3}{}^{37}$Cl which we provide as supplementary material.~\cite{EPAPSCH3CL} We recommend the CBS-35$^{{\,}\mathrm{HL}}$ and CBS-37$^{{\,}\mathrm{HL}}$ PESs for future use.
 
\subsection{Equilibrium Geometry of CH$_3$Cl}

 The equilibrium geometry of methyl chloride determined empirically by Jensen et al.~\cite{81JeBrGu.CH3Cl} has often served as reference. However the reliability of the axial rotational constants used in their analysis has been questioned.~\cite{97DeWlRu.CH3Cl} The C-H bond length reported in Ref.~\onlinecite{81JeBrGu.CH3Cl} also appears too large to be consistent with \textit{ab initio} calculations, and also with the isolated C-H bond stretching frequency.~\cite{94DeWlxx.CH3Cl} A combined empirical and \textit{ab initio} structure has later been determined based on $^{12}$CH$_{3}{}^{35}$Cl, $^{12}$CH$_{3}{}^{37}$Cl, $^{12}$CD$_{3}{}^{35}$Cl and $^{12}$CD$_{3}{}^{37}$Cl experimental data.~\cite{97DeWlRu.CH3Cl} We compare against this as well as another high-level \textit{ab initio} study.~\cite{03DeMaBo.CH3Cl}

 The equilibrium structural parameters calculated from the CBS-(35/37)$^{\,\mathrm{HL}}$ PES and the VQZ-F12 PES are listed in Table~\ref{tab:eq_ref}. The CBS-(35/37)$^{\,\mathrm{HL}}$ bond lengths are shorter than the VQZ-F12 values, which is to be expected due to the inclusion of core-valence electron correlation.~\cite{Coriani:2005} There is good agreement with the values from Refs.~\onlinecite{97DeWlRu.CH3Cl} and \onlinecite{03DeMaBo.CH3Cl}. The largest discrepancy concerns the bond angle determined in Ref.~\onlinecite{97DeWlRu.CH3Cl} which is around $0.3$ degrees larger than all \textit{ab initio} computed values.
 
\begin{table}
\tabcolsep=5pt
\caption{\label{tab:eq_ref}Equilibrium structural parameters of CH$_3$Cl}
\begin{center}
\begin{tabular}{l c c c}
\hline\hline
& $r$(C-Cl)/$\mathrm{\AA}$ & $r$(C-H)/$\mathrm{\AA}$ & $\beta$(HCCl)/deg  \\[0.5mm]
\hline  \\[-2.5mm]
CBS-(35/37)$^{\,\mathrm{HL}}$ & 1.7777 & 1.0834 & 108.38 \\
VQZ-F12 & 1.7805 & 1.0849  & 108.39 \\ 
Ref.~\onlinecite{97DeWlRu.CH3Cl}$\,^a$ & 1.7768 & 1.0842 & 108.72 \\
Ref.~\onlinecite{03DeMaBo.CH3Cl}$\,^b$ & 1.7772 & 1.0838 & 108.45 \\
\hline\hline
\end{tabular}
\end{center}
$^a$ Value determined from empirical data and CCSD(T) calculations.\\[-2mm]
$^b$ CCSD(T)(\textit{fc})/cc-pV(Q,5)Z + MP2(\textit{ae})/cc-pwCVQZ - MP2(\textit{fc})/cc-pwCVQZ.
\end{table}
 
 For further validation we studied the pure rotational spectrum as rotational energies are highly dependent on the molecular geometry through the moments of inertia. In Table~\ref{tab:rotational} we present the calculated $J\leq5$ rotational energies in the ground vibrational state for CH$_{3}{}^{35}$Cl using the CBS-35$^{\,\mathrm{HL}}$ PES. The computed values reproduce the experimental levels with a rms error of $0.0018{\,}$cm$^{-1}$. The CBS-(35/37)$^{\,\mathrm{HL}}$ \textit{ab initio} structural parameters reported in Table~\ref{tab:eq_ref} can thus be regarded as reliable, and we expect the true equilibrium geometry of methyl chloride to be close to these values.
 
\begin{table}[!h]
\tabcolsep=5pt
\caption{\label{tab:rotational} Comparison of calculated and experimental $J\leq5$ pure rotational term values (in cm$^{-1}$) for CH$_{3}{}^{35}$Cl. The observed ground state energy levels are from Ref.~\onlinecite{05NiChxx.CH3Cl}.}
\begin{center}
	\begin{tabular}{cccrrr}
	\hline\hline
 	 $J$ & $K$ & Sym. &  Exp. & CBS-35$^{{\,}\mathrm{HL}}$ & Obs$-$calc \\ 
	\hline	
	0 & 0 & $A_1$ &   0.0000 &   0.0000 &  0.0000 \\
	1 & 0 & $A_2$ &   0.8868 &   0.8868 &  0.0000 \\
	1 & 1 & $E$   &   5.6486 &   5.6489 & -0.0003 \\
	2 & 0 & $A_1$ &   2.6604 &   2.6603 &  0.0001 \\
	2 & 1 & $E$   &   7.4222 &   7.4223 & -0.0001 \\
	2 & 2 & $E$   &  21.7067 &  21.7075 & -0.0008 \\
	3 & 0 & $A_2$ &   5.3208 &   5.3205 &  0.0003 \\
	3 & 1 & $E$   &  10.0825 &  10.0826 & -0.0001 \\
	3 & 2 & $E$   &  24.3668 &  24.3676 & -0.0008 \\
	3 & 3 & $A_1$ &  48.1707 &  48.1727 & -0.0020 \\
	3 & 3 & $A_2$ &  48.1707 &  48.1727 & -0.0020 \\
	4 & 0 & $A_1$ &   8.8678 &   8.8675 &  0.0003 \\
	4 & 1 & $E$   &  13.6295 &  13.6294 &  0.0001 \\
	4 & 2 & $E$   &  27.9137 &  27.9143 & -0.0006 \\
	4 & 3 & $A_1$ &  51.7173 &  51.7191 & -0.0018 \\
	4 & 3 & $A_2$ &  51.7173 &  51.7191 & -0.0018 \\
	4 & 4 & $E$   &  85.0354 &  85.0389 & -0.0035 \\
	5 & 0 & $A_2$ &  13.3015 &  13.3010 &  0.0005 \\
    5 & 1 & $E$   &  18.0632 &  18.0629 &  0.0003 \\
    5 & 2 & $E$   &  32.3472 &  32.3476 & -0.0004 \\
    5 & 3 & $A_1$ &  56.1505 &  56.1521 & -0.0016 \\
    5 & 3 & $A_2$ &  56.1505 &  56.1521 & -0.0016 \\
    5 & 4 & $E$   &  89.4681 &  89.4714 & -0.0033 \\
    5 & 5 & $E$   & 132.2931 & 132.2985 & -0.0054 \\
    \hline\hline
    \end{tabular}
\end{center}
\end{table}
 
\section{Conclusions}
\label{sec:conc}

 Using state-of-the-art electronic structure calculations, we have generated two global \textit{ab initio} PESs for the two main isotopologues of methyl chloride. We believe that the accuracy achieved by these PESs is at the limit of what is currently possible using solely \textit{ab initio} methods, and that it would be extremely challenging to go beyond this without empirical refinement of the respective PESs. Considering that the PESs are purely \textit{ab initio} constructs, the computed vibrational wavenumbers show remarkable accuracy when compared with experiment for both isotopologues. It is evident that higher-level energy corrections and an extrapolation to the complete basis set limit should be included to obtain accurate theoretical vibrational energies for CH$_3$Cl. The same applies to the determination of equilibrium structural parameters.
 
 For the requirements of high-resolution spectroscopy, the \textit{ab initio} surfaces presented here will no doubt need to be refined to empirical data.~\cite{YuBaTe11.NH3} The resulting ``spectroscopic PESs'' can then be used to achieve unprecedented accuracy in the simulation of rotation-vibration spectra, and it is at this stage that the predictive power of the variational approach is fully realised. The PESs presented in this work provide an excellent starting point for this procedure. Once the refinement is complete, a comprehensive rovibration line list applicable for elevated temperatures will be generated for methyl chloride as part of the ExoMol project.~\cite{ExoMol2012}

\begin{acknowledgments}
This work was supported by ERC Advanced Investigator Project 267219, and FP7-MC-IEF project 629237. A.O. is grateful to UCL for a studentship under their Impact Student Scheme and thanks J\"{u}rgen Breidung for valuable discussions.
\end{acknowledgments}

%

\end{document}